\documentclass[twocolumn,showpacs,showkeys]{revtex4}
\usepackage{epsfig}
\usepackage{graphicx}
\input{psfig.sty}

\newcommand{\bastar}{\begin{eqnarray*}}
\newcommand{\eastar}{\end{eqnarray*}}
\newskip\humongous \humongous=0pt plus 1000pt minus 1000pt

\newif\ifdtup

\relax
\newcommand{\be}{\begin{equation}}
\newcommand{\ee}{\end{equation}}
\newcommand{\bea}{\begin{eqnarray}}
\newcommand{\eea}{\end{eqnarray}}
\newcommand{\X}{{\vec X}}
\newcommand{\pro}{\partial}
\newcommand{\n}{\hat n}
\newcommand{\oneg}{\displaystyle\frac{1}{g}}

\newcommand{\D}{{\hat D}}

\newcommand{\valpha}{{\vec \alpha}}

\newcommand{\hn}{{\hat n}}
\newcommand{\hD}{{\hat D}}
\newcommand{\dfrac}{\displaystyle\frac}
\newcommand{\ba}{\begin{array}}
\newcommand{\ea}{\end{array}}

\newcommand{\nn}{\nonumber}

\begin{document}
\title{Stability of Monopole Condensation in SU(2) QCD}
\bigskip

\author{Y. M. Cho}
\email{ymcho@yongmin.snu.ac.kr}
\author{M. L. Walker}
\email{mlwalker@phya.snu.ac.kr}
\affiliation{School of Physics, College of Natural Sciences,
Seoul National University, Seoul 151-747, Korea}
\begin{abstract}
We propose to resolve the controversy regarding the stability
of the monopole condensation in QCD by calculating the imaginary part
of the one-loop effective action perturbatively. We calculate
the imaginary part perturbatively to the order $g^2$ with two
different methods, with Fyenman diagram and with Schwinger's method.
Our result shows that with the magnetic background
the effective action has no imaginary part, but with the electric
background it acquires a negative imaginary part. This strongly
indicates a stable monopole condensation in QCD.
\end{abstract}
\maketitle

One of the most outstanding problems
in theoretical physics is the confinement problem in
quantum chromodynamics (QCD). It has
long been argued that monopole condensation can explain the
confinement of color through the dual Meissner effect
\cite{nambu,cho80}.
Indeed, if one assumes monopole condensation, one can easily argue
that the ensuing dual Meissner effect guarantees the confinement.
A satisfactory proof of the desired monopole condensation in QCD,
however, has been very elusive \cite{savv,niel,ditt,cons}.

There have been many attempts to demonstrate monopole condensation
in QCD from the one-loop effective action.
Savvidy first calculated the effective action
of $SU(2)$ QCD in the presence of an {\it ad hoc}
color magnetic background, and discovered an encouraging
evidence of magnetic condensation as a non-trivial vacuum
of QCD \cite{savv}. But soon Nielsen and Olesen repeated the
calculation and found that
the effective action with the magnetic background generates
an imaginary part which makes the vacuum unstable \cite{niel}.
The origin of this instability of the ``Savvidy-Nielsen-Olesen (SNO)
vacuum'' was traced back to the fact that the quantum fluctuation
around the magnetic background contains tachyonic modes
which destabilizes the magnetic condensation. This instability of
SNO vacuum has been re-examined by many authors, but never been 
seriously challenged \cite{ditt,cons}. 
This has created an unfortunate impression that
it might be impossible to establish the monopole condensation
with the one-loop effective action of QCD.

Recently, however, there has been a new calculation of
the one-loop effective action of QCD
with a gauge independent separation of the classical background
from the quantum field \cite{cho02,cho99}.
Remarkably, in this calculation the effective action has been shown to
produce no imaginary part in the presence of
the non-Abelian monopole background, but a negative imaginary part
in the presence of the pure color electric background.
A major difference between this and the old calculations is
the infra-red regularization. While the old calculations used the
$\zeta$-function regularization, the new calculation adopted
an infra-red regularization which respects causality \cite{cho02,cho99}.

The new result sharply contradicts with the old result,
and has brought the controversy on the
stability of the monopole condensation back to life.
It is therefore imperative that we look for an
independent evidence of the monopole condensation which
could settle the controversy once and for all.

A remarkable point of gauge theories is that in the massless limit
the imaginary part of the effective action
becomes proportional to $g^2$, where $g$ is the gauge coupling
constant. This has been shown to be true
in both QCD \cite{niel,ditt,cho02,cho99} and massless
QED \cite{cho01prl,cho01}. This suggests that one could calculate
the imaginary part of the effective action perturbatively
to the order $g^2$, and check the presence (or absence) of
the imaginary part with the perturbative method \cite{sch}.
{\it The purpose of this Letter is to calculate
the imaginary part of the one-loop effective action of $SU(2)$ QCD
perturbatively to the order $g^2$.
Our result shows that the effective action has no imaginary part
in the presence of the monopole background but has a negative
imaginary part in the presence of the color electric background}.
This endorses the new result based on the infra-red regularization
by causality \cite{cho02,cho99}, which strongly supports
the stability of the monopole condensation in QCD.

We start by reviewing the Abelian formalism
of QCD \cite{cho80,cho00}, which we need to make the comparison
between QCD and massless QED more transparent.
For simplicity we concentrate on $SU(2)$ QCD in this paper.
We introduce a gauge-covariant unit isovector $\n$
which selects the color direction, and decompose
the gauge potential into the binding potential $\hat A_\mu$
and the valence potential $\vec X_\mu$ \cite{cho80,cho00},
\bea
\label{cdec}
& \vec{A}_\mu =A_\mu \n - \oneg \n\times\pro_\mu\n+\X_\mu\nonumber
         = \hat A_\mu + \X_\mu, \nn\\
& (A_\mu = \n\cdot \vec A_\mu,~ \n^2 =1,~ \hat{n}\cdot\vec{X}_\mu=0),
\eea
where $A_\mu$ is the ``electric'' potential.
Notice that $\hat A_\mu$ is precisely
the connection which leaves $\n$ invariant under parallel transport,
\bea
\D_\mu \n = \pro_\mu \n + g {\hat A}_\mu \times \n = 0.
\eea
Under the infinitesimal gauge transformation
one has
\bea
&&\delta A_\mu = \oneg \n \cdot \pro_\mu \valpha,\,\,\,\
\delta \hat A_\mu = \oneg \D_\mu \valpha  ,  \nn \\
&&\hspace{1.2cm}\delta \X_\mu = - \valpha \times \X_\mu  .
\label{gt}
\eea
This tells that
$\hat A_\mu$ by itself describes an $SU(2)$
connection which enjoys the full $SU(2)$ gauge degrees of
freedom. Besides, it has a dual structure,
\begin{eqnarray}
& \hat{F}_{\mu\nu} = (F_{\mu\nu}+ H_{\mu\nu})\hat{n}\mbox{,}
\nonumber \\
& F_{\mu\nu} = \partial_\mu A_{\nu}-\partial_{\nu}A_\mu \mbox{,}
\nonumber \\
& H_{\mu\nu} = -\dfrac{1}{g} \hat{n}\cdot(\partial_\mu
\hat{n}\times\partial_\nu\hat{n})
= \partial_\mu \tilde C_\nu-\partial_\nu \tilde C_\mu,
\end{eqnarray}
where $\tilde C_\mu$ is the ``magnetic'' potential of
the non-Abelian monopoles \cite{cho80,cho80prl}.
Furthermore, (\ref{gt}) shows that $\vec X_\mu$ forms a
gauge covariant vector field.
But most importantly, the decomposition (\ref{cdec}) is
gauge independent. Once $\hat n$ is given, the decomposition
uniquely defines $\hat A_\mu$ and $\vec X_\mu$, independent of
the choice of gauge \cite{cho80,cho00}.

To Abelianize QCD let
$(\hat n_1,~\hat n_2,~\hat n)$
be a right-handed orthonormal basis in $SU(2)$ space and let
\bea
B_\mu = A_\mu + \tilde C_\mu, ~~~~X_\mu = \dfrac{1}{\sqrt{2}}
( \hat{n}_1 + i \hat{n}_2 ) \cdot \vec{X}_\mu.
\eea
With this the Yang-Mills Lagrangian is written as
\begin{eqnarray} \label{aqcd}
&{\cal L}=-\dfrac{1}{4} G_{\mu\nu}^2
-\dfrac{1}{2}|\hat{D}_\mu{X}_\nu-\hat{D}_\nu{X}_\mu|^2
+ ig G_{\mu\nu} X_\mu^* X_\nu \nn\\
&-\dfrac{1}{2} g^2 \Big[(X_\mu^*X_\mu)^2-(X_\mu^*)^2 (X_\nu)^2 \Big],
\end{eqnarray}
where now
\bea
G_{\mu\nu} = \partial_\mu B_\nu - \partial_\nu B_\mu,
~~~~~\hat{D}_\mu{X}_\nu = (\partial_\mu + ig B_\mu) X_\nu.\nonumber
\eea
Clearly this describes an Abelian gauge theory coupled to
the charged vector field $X_\mu$, except
that here the Abelian potential $B_\mu$ actually has
a dual structure. It contains both electric and magnetic potentials
\cite{cho80,cho00}.

Notice that this Abelianization is
gauge independent, because here we have never fixed
the gauge to obtain the Abelian formalism.
To see how the non-Abelian symmetry is retained in this
Abelian formalism let
\bea
&\vec \alpha = \alpha_1\hn_1 + \alpha_2\hn_2 + \theta\hat n,
~~~\alpha = \dfrac{1}{\sqrt 2} (\alpha_1 + i \alpha_2), \nn\\
&C_\mu = \dfrac{1}{\sqrt 2} (\hn_1 + i \hn_2)
\cdot (\dfrac {1}{g} \hn \times \partial_\mu \hn).
\eea
Then the Lagrangian (\ref{aqcd}) is invariant not only under
the active gauge transformation (\ref{gt})
\bea \label{actgt}
&\delta A_\mu = \dfrac{1}{g} \partial_\mu \theta -
i (C_\mu^* \alpha - C_\mu \alpha^*),
~~~&\delta \tilde C_\mu = - \delta A_\mu, \nn\\
&\delta X_\mu = 0,
\eea
but also under the following passive gauge transformation
\bea \label{pasgt}
&\delta A_\mu = \dfrac{1}{g} \partial_\mu \theta -
i (X_\mu^* \alpha - X_\mu \alpha^*), ~~~&\delta \tilde C_\mu = 0, \nn\\
&\delta X_\mu = \oneg \hD_\mu \alpha - i \theta X_\mu.
\eea
This tells that the Abelian formalism
actually has enlarged (both the
active and passive) gauge symmetries \cite{cho80,cho00}.

We can calculate the effective action of QCD
with the Abelian formalism.
We first identify $B_\mu$ as the classical background
and $X_\mu$ as the fluctuating quantum
part, and fix the gauge of the quantum field by
imposing the gauge condition,
\bea
\hat D_\mu X_\mu =0, ~~~~~{\cal L}_{gf} =- \dfrac{1}{\xi}
|{\hat D}_\mu X_\mu|^2.
\eea
With the gauge fixing we have two functional
determinants which contribute to
the effective action, the valence gluon and
the ghost determinant. So we have
\bea
&\Delta S = i\ln {\rm Det} \Big[(-\hD^2+2a)(-\hD^2-2a)  \nn\\
&(-\hD^2-2ib)(-\hD^2+2ib) \Big] - 2i\ln {\rm Det}(-\hD^2),
\eea
where
\bea
a &=& \dfrac{g}{2} \sqrt {\sqrt {G^4 + (G \tilde G)^2} + G^2},
\nn\\
b &=& \dfrac{g}{2} \sqrt {\sqrt {G^4 + (G \tilde G)^2} - G^2}. \nn
\eea
One can evaluate the functional determinants
and find \cite{ditt,cho02,cho99}
\bea
&\Delta {\cal L} =  \dfrac{1}{16 \pi^2}  \int_{0}^{\infty}
\dfrac{dt}{t^3} \dfrac{ a b t^2 / \mu^4}{\sinh (at/\mu^2)
\sin (bt/\mu^2)} \nn\\
&\times \Big[\exp(-2at/\mu^2)+\exp(2at/\mu^2)+\exp(2ibt/\mu^2) \nn\\
&+\exp(-2ibt/\mu^2)-2 \Big],
\label{ea}
\eea
where $\mu$ is a dimensional parameter. Notice that the monopole
background is described by $a\neq0$ and $b=0$.

The evaluation of the integral (\ref{ea})
has been notoriously difficult. In fact,
even in the case of much simpler QED, the calculation of the
effective action has been completed only recently \cite{cho01prl,cho01},
fifty years after the seminal work of Schwinger \cite{schw}.
For $b=0$ the old calculations of the integral (\ref{ea})
using the $\zeta$-function regularization
gives \cite{savv,niel,ditt}
\bea
{\cal L}_{eff}{\Big|}_{SNO} &=& -\dfrac{a^2}{2g^2} -\dfrac{11 a^2}{48\pi^2}
\ln \dfrac{a}{\mu^2} \nn\\
&+& i\dfrac{a^2}{8\pi}~~~~~(b=0).
\label{snoea}
\eea
Notice that the real part of this SNO effective Lagrangian
generates a magnetic condensation. Unfortunately this
magnetic condensation is destablized by the imaginary part
of the Lagrangian with pair-creation of gluons.

But we can integrate (\ref{ea}) with the infra-red regularization
by causality, and obtain \cite{cho02,cho99} 
\bea 
&{\cal L}_{eff} = \left\{\begin{array}{ll}-\dfrac{a^2}{2g^2} 
-\dfrac{11a^2}{48\pi^2}(\ln \dfrac{a}{\mu^2}-c )&~~ (b=0), \\
\dfrac{b^2}{2g^2} +\dfrac{11b^2}{48\pi^2} (\ln \dfrac{b}{\mu^2}-c)\\
-i\dfrac{11b^2}{96\pi} &~~ (a=0), \end{array}\right. 
\label{choea} 
\eea where $c$ is a subtraction-dependent constant. Observe that
this effective Lagrangian has no omaginary part when $b=0$.
Moreover it has a remarkable symmetry called the duality. It is
invariant under the dual transformation $a \rightarrow - ib$, and
$b \rightarrow ia$. This duality was recently established as a
fundamental symmetry of the effective action of gauge theory, both
Abelian and non-Abelian \cite{cho02,cho01}. The duality provides a
very useful tool to check the self-consistency of the effective
action.

Clearly the effective Lagrangian (\ref{choea}) provides the following
effective potential in the presence of the magnetic background
\bea
V=\dfrac12 \dfrac{a^2}{g^2}
\Big[1+\dfrac{11 g^2}{24 \pi^2}(\ln\dfrac{a}{\mu^2}-c)\Big],
\label{epot}
\eea
which generates a non-trivial local minimum at
\bea \label{eq:minimum}
<a>=\dfrac{\mu^2}{\sqrt 2} \exp\Big(-\dfrac{24\pi^2}{11g^2}+
c-\dfrac12\Big).
\eea
This is nothing but the desired monopole condensation \cite{cho02,cho99}.
The effective potential (\ref{epot}) is
shown in Fig. 1.

\begin{figure}
\begin{center}
\psfig{figure=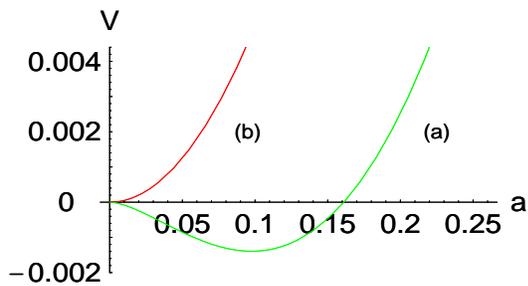, height = 3.8 cm, width = 7 cm}
\end{center}
\caption{\label{Fig. 1} The effective potential of SU(2) QCD in
the pure magnetic background. Here (a) is the effective potential
and (b) is the classical potential.}
\end{figure}

Apparently the difference between two effective Lagrangians
(\ref{snoea}) and (\ref{choea}) follows from different infra-red
regularizations. In (\ref{snoea}) it was the $\zeta$-function
regularization, but in (\ref{choea}) it was the infra-red
regularization by causality. Since the $\zeta$-function
regularization is such a well established regularization
we can not easily dismiss the instability of the monopole
condensation. So we have to know which
regularization is the correct one, and why. We need an independent
method which can settle this controversy.

Fortunately we can settle this controversy with a perturbative
method, because the imaginary part of the effective
action is of the order of $g^2$.
The idea that one can actually settle this controversy with
a perturbative method was first proposed
by Schanbacher \cite{sch}, but this idea has never been tested by actual
calculation before. So we first demonstrate that in massless
QED one can indeed calculate the imaginary part perturbatively,
and apply the perturbative method to obtain the imaginary part
of the QCD effective action.
To do this we review the Schwinger's
perturbative calculation of the QED effective action. In QED
Schwinger has obtained the following effective action
perturbatively to the order $e^2$ \cite{schw}
\bea
&\Delta S_{QED}=\dfrac{e^2}{16 \pi^2} \int d^4p
F_{\mu\nu}(p)F_{\mu\nu}(-p) \nn\\
&\times \dfrac{}{}\int_{0}^{1} dv \dfrac{v^2 (1- v^2/3)}{(1- v^2)
+ 4m^2/p^2},
\label{qedea}
\eea
where $m$ is the electron mass. From this he observed that
when $-p^2>4m^2$ the integrand develops a pole at
$v^2=1+4m^2/p^2$ which generates an imaginary part, and
explained how to calculate the imaginary part
of the effective action. But notice that in the massless limit,
the pole moves to $v=1$. In this case the pole contribution
to the imaginary part is reduced by a half, and we obtain
\bea
Im ~{\cal L}_{QED} {\Big |}_{m=0} = \left\{{~0
~~~~~~~~~~~~~~ b=0,
\atop \dfrac{b^2}{48 \pi} ~~~~~~~~~~~a=0.}\right.
\eea
{\it Remarkably, this is exactly what we obtain
from the non-perturbative effective action in
the massless limit} \cite{cho01prl,cho01}.
This confirms that in massless QED, one can calculate the
imaginary part of the effective action either perturbatively or
non-perturbatively, with identical results.

Now we repeat the perturbative calculation for QCD.
We can do this either by calculating
the one-loop Feynman diagrams directly,
or by evaluating the integral (\ref{ea}) perturbatively to
the order $g^2$. We start with the Feynman diagrams.
For an arbitrary background $B_\mu$ there are four
Feynman diagrams that contribute to the order $g^2$ which are shown
in Fig. 2. Notice that the tadpole diagrams contain a quadratic
divergence which does not appear in the final result.

\begin{figure}
\begin{center}
\psfig{figure=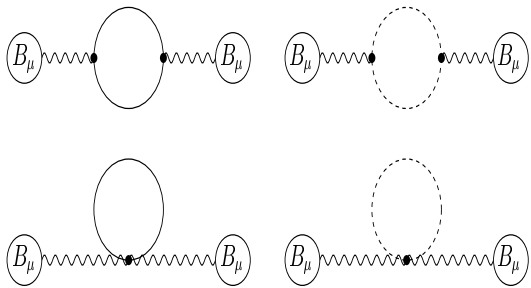, height = 3.8 cm, width = 7 cm}
\end{center}
\caption{\label{Fig. 2} The Feynman diagrams that contribute to the
effective action at $g^2$ order. Here the straight line and
the dotted line represent the
valence gluon and the ghost, respectively.}
\end{figure}

The sum of these diagrams (in the Feynman gauge with dimensional
regularization) gives us \cite{pesk}
\bea \label{peafeyn}
&\Delta S = -\dfrac{11g^2}{96 \pi^2} \int d^4p
G_{\mu\nu}(p)G_{\mu\nu}(-p) \nn\\
&\times \left[\mbox{ln}
\left(\dfrac{p^2}{\mu^2}\right) + C_1 \right],
\eea
where $C_1$ is a regularization-dependent constant.
Clearly the imaginary part could only
arise from the term $\mbox{ln} (p^2/\mu^2)$, so that
for a space-like $p^2$ (with $\mu^2>0$) the effective action
has no imaginary part. However, since a space-like
$p^2$ corresponds to a magnetic background, we find that
the magnetic condensation generates no imaginary part,
at least at the order $g^2$. To evaluate the imaginary part
for a constant electric background we have to make the analytic
continuation of (\ref{peafeyn}) to a time-like $p^2$,
because the electric background corresponds to a time-like $p^2$.
In this case the causality (with the familiar Feynman
prescription $p^2 \rightarrow
p^2-i \epsilon$) dictates us to have
\bea
&\mbox{ln} \Big(\dfrac{p^2}{\mu^2}\Big) \rightarrow 
\lim_{\epsilon \to0} \mbox{ln} \Big(\dfrac{p^2-i\epsilon}{\mu^2} \Big) \nn\\
&=\mbox{ln} \Big(\dfrac{|p^2|}{\mu^2}\Big)- i \dfrac{\pi}{2}
~~~~~(p^2<0),
\label{causal}
\eea
so that we obtain
\bea Im ~\Delta {\cal L} = \left\{{~~~~0~~~~~~~~~~~~(b=0),
\atop -\dfrac{11 b^2}{96 \pi} ~~~~~~~~~(a=0).}\right.
\label{imea}
\eea
This allows us to conclude that the result (\ref{choea}) is
indeed endorsed by
the Feynman diagram calculation.

To remove any lingering doubt about (\ref{choea}) we now make the
perturbative calculation of the integral (\ref{ea}) to the order
$g^2$ with the Schwinger's method, and find \cite{hon}
\bea
&\Delta S = -\dfrac{g^2}{8\pi^2}
\int d^4p G_{\mu \nu}(p) G_{\mu \nu}(-p) \Sigma (p), \nn\\
&\Sigma (p) = \dfrac{}{}\int_{0}^{1} dv (1-\dfrac{v^2}{4})
\dfrac{}{}\int_{0}^{\infty}\dfrac{dt}{t}
\exp[-\dfrac{p^2}{4}(1-v^2) t] \nn\\
&= 2\dfrac{}{}\int_{0}^{1} dv \dfrac{v^2(1-v^2/12)}{1-v^2} + C_2,
\label{peasch}
\eea
where $C_2$ is a regularization-dependent constants.
Now, it is straightforward to evaluate the imaginary part of $\Delta S$.
Comparing this with Schwinger's result (\ref{qedea}) for
the massless QED we again reproduce (\ref{imea}),
after the proper charge and wave function
renormalization.

Furthermore, from the definition of the exponential
integral function \cite{table}
\bea
&Ei (-z) = - \dfrac{}{} \int_{z}^{\infty} \dfrac{d\tau}{\tau}
\exp (-\tau) = \gamma + \mbox{ln} z \nn\\
&+ \dfrac{}{} \int_{0}^{z} \dfrac{d\tau}{\tau}
\left[\exp (-\tau)-1\right]~~~({\rm Re}~z > 0),
\eea
we can express $\Sigma (p)$ as
\bea
&\Sigma (p) = - \dfrac{}{}\lim_{\epsilon\to0} \int_{0}^{1} dv
(1-\dfrac{v^2}{4}) \left[ \gamma + \mbox{ln} \Big(\dfrac{p^2}{4 \mu^2}(1-v^2)
\epsilon \Big) \right] \nn\\
&= \dfrac{11}{12}\left[\mbox{ln}\left(\dfrac{p^2}{\mu^2}\right) + C_3 \right],
\label{sigma}
\eea
where $C_3$ is another
regularization-dependent constant.
This tells that (\ref{peasch}) is identical to (\ref{peafeyn}).
This is the reason why the perturbative calculation by Feynman diagrams
and by Schwinger's method produce the same result.
{\it This strongly indicates that the monopole condensation
indeed describes a stable vacuum, but the electric background creates
the pair-annihilation of the valence gluons in
$SU(2)$ QCD} \cite{cho02,cho99}.

It is striking that both the infra-red regularization by causality and
the perturbative method endorse the stability of the monopole
condensation. But in retrospect one should not be surprised by this.
Remember that the instability of SNO vacuum originates
from the tachyonic modes which violate the causality.
In physics the appearence of tachyonic modes has always
implied that something is wrong in the formalism, and one
corrects this defect by introducing a physical condition
which can exclude the tachyonic modes. This is what happens in
string theory and in spontaneously broken gauge theory.
And in QCD obviously the causality is what we need
to exclude the unphysical tachyonic modes. So it is
natural that both the perturbative method and the infra-red
regularization by causality ensure the stability of
the monopole condensation in QCD, because both are based on
the causality.

This does not mean that the $\zeta$-function regularization
has any intrinsic defect. We emphasize that the problem
is the incorrect inclusion of the unphysical tachyonic modes
in the integral (\ref{ea}), not the $\zeta$-function regularization.
The $\zeta$-function regularization is simply too honest to
remove the tachyonic modes.

In this paper we have neglected the quarks. We simply remark
that the quarks, just as in asymptotic freedom \cite{wil},
tend to destabilize the monopole condensation. In fact the stability
puts exactly the same constraint on the number of quarks as
the asymptotic freedom \cite{cho99,cho1}.

Recently the monopole condensation has been establshed in
a supersymmetric generalization of QCD \cite {witt}.
Our analysis tells that one can establsh the magnetic
condensation within the framework of QCD, with the existing
principles of quantum field theory.
It is truly remarkable (and surprising) that the principles of
quantum field theory allow us to demonstrate the magnetic
condensation, and by implication the confinement of
color, within the framework of QCD. This should be interpreted as a
most spectacular triumph of quantum field theory itself.

We conclude with the following remarks: \\
1) We emphasize that the above perturbative calculation
of the  imaginary part of the one-loop effective action
was possible because in QCD (and in massless QED)
the imaginary part of the one-loop effective action is
of the order $g^2$. This assures us that one can make
a perturbative expansion for the imaginary part of the effective action.
For massive QED, for example, this calculation does not make sense because
the imaginary (as well as the real) part of the effective action
simply does {\it not} allow a convergent
perturbative expansion \cite{cho01prl,cho01}. The same argument
applies to the real part of QCD. Only
for the imaginary part of the massless gauge theories one can make
sense out of the perturbative calculation. \\
2) One might worry about the negative signature of the imaginary part
in the QCD effective action. To understand the origin
of this, compare QCD with
massless QED. The difference between the two is that
in QED we have the electron loop, but in QCD we have the valence gluon loop.
Obviously they have the opposite statistics (aside from the different
kinematic factors which do not change the signature).
This is the reason for the negative signature \cite{cho02,cho99}.
This implies that the electric background generates
pair-annihilation (not pair-creation) of gluons in QCD.
And this is what we need to explain the infra-red slavery,
because the pair-annihilation implies the anti-screeing of color.
This tells that the negative signature is consistent with the
confinement of color.  \\
3) In this paper we have considered only the pure magnetic
or pure electric background, so
the above result guarantees only the stability of the
monopole condensation. To show that this
is the true vacuum of QCD, we must calculate the effective action
with an arbitrary background. Fortunately, one
can actually do this with an arbitrary
constant background, and show that indeed the monopole condensation
becomes the true vacuum of $SU(2)$ QCD, at least at one-loop
level \cite{cho99,cho1}.

The details of our analysis will
be published elsewhere \cite{cho2}.

One of the authors (YMC) thanks Professor S. Adler, Professor F. Dyson,
and Professor C. N. Yang for the illuminating discussions.
The work is supported in part by the Basic Research Program
(Grant R02-2003-000-10043-0) of
Korea Science and Engineering Foundation and by
the BK21 project of Ministry of Education.

\end{document}